\documentstyle[12pt,psfig]{article}
\begin{document}


%
%
\def\cm2{${\rm cm^{-2}}$}
\def\nhi{$N_{HI}$}
\def\ly{Lyman$-\alpha$} 
\def\h1a{HI 21 cm-line absorption}
\def\etal{et al. }
\def\kms{kms$^{-1}$}
\title{GMRT detection of HI 21 cm associated absorption \\
towards the z=1.2 red quasar 3C 190}

\author{C. H. Ishwara-Chandra$^1$\thanks{e-mail:ishwar@ncra.tifr.res.in}, 
~K. S. Dwarakanath$^2$\thanks{e-mail:dwaraka@rri.res.in}\\
and  K. R. Anantharamaiah$^2$ \thanks{Deceased}\\
{\normalsize $^1$National Center for Radio Astrophysics, TIFR,} \\
{\normalsize Post Bag 3, Ganeshkhind, Pune 411 007, India.}\\
{\normalsize $^2$Raman Research Institute, Sadashivanagar, Bangalore - 560 080, India.}
}

\date{}

\maketitle

\begin{abstract}

We report the GMRT detection of associated HI 21 cm-line absorption in
the z=1.1946 red quasar 3C 190. Most of the absorption is blue-shifted
with respect to the systemic redshift. The absorption, at $\sim$ 647.7
MHz, is broad and complex, spanning a velocity width of $\sim$ 600 \kms.
Since the core is self-absorbed at this frequency, the absorption is most
likely towards the hotspots. Comparison of the radio and deep optical
images reveal linear filaments in the optical which overlap with the
brighter radio jet towards the south-west. We therefore suggest that 
most of the \h1a could be occurring in the atomic gas shocked by the south-west jet.

\end{abstract}

\noindent {\bf Keywords:} galaxies: active - quasars: absorption-lines - radio lines: galaxies
quasars:individual(3c 190)

\section{Introduction}

It was suggested more than three decades ago that the study of HI
21 cm-line absorption at high redshifts would provide interesting and
important information regarding the distribution and kinematics of neutral
hydrogen at earlier epochs (Bahcall and Ekers 1969).  The central regions
of active galactic nuclei (AGN) are important in understanding many
aspects of AGN phenomena like the fueling on to the accretion disk and the
obscuration of the nuclei as proposed in the unified scheme. One way to
study the central region in AGNs is to search for \h1a at the redshift of
the AGN host galaxy against the radio source. The absorption can be due
to a variety of phenomena like tori, outflows, the inter stellar medium
(ISM), cold clouds and cold gas stirred up during merging of neighbors
with the host galaxy (cf. Morganti \etal 2001).

The red quasars have generated interest particularly since the work of
Webster \etal (1995) claiming that a large fraction ($\sim$ 80\,\%) of
quasars could be missed from optical surveys due to dust extinction. It
is believed that in this class of quasars, the extinction due to dust is
higher (Webster \etal 1995); however, there are claims that red quasars
are not necessarily dusty (eg: Benn \etal 1998).  If the red quasars are
dusty, then the number of high column density absorption-line systems seen
towards optically selected quasars will also be an underestimate. Carilli
\etal (1998; hereinafter C98) have searched for \h1a in a radio-loud
red quasar subsample at moderate redshifts (z $\sim$ 0.7) to address
whether the 'red' AGNs are intrinsically red or reddened by dust. In
their sample of five red quasars, four showed significant HI absorption,
suggesting the presence of large amounts of gas and associated dust. Even
though the number is small, this study showed that the success rate for
detecting \h1a in radio-loud red quasars is higher (80\,\%) compared to
an optically selected sample of radio-loud quasars with Mg II or Lyman
alpha absorption (C98, Carilli \etal 1999).  This opens up a new class
of objects which can be studied in 21-cm absorption at higher redshifts,
where the optically selected sample has strong bias against high column
density systems with high dust-to-gas ratios (Fall and Pei 1993 1995).

We have started a program to search for \h1a in a sample of high redshift
(z $>$ 1) radio-loud 'red' quasars and galaxies with the Giant Meterwave
Radio Telescope (GMRT, Swarup \etal 1991). At higher redshifts, the
optical/IR morphologies of the host galaxies tend to be more complex,
possibly due to ongoing merging activity leading to the formation of
massive ellipticals at later epochs (Chambers and Miley 1990). Such
activity (i) might enhance the probability of detecting \h1a (eg:
Carilli and van Gorkom 1992) and (ii) could also contribute to reddening.

3C 190 is a compact steep spectrum (CSS) radio source. Its largest
angular size is 4.0 arcsec and corresponding linear size is 33 kpc
(assuming H$_\circ$=71\kms, $\Omega_m$ = 0.27 and $\Omega_{vac}$ = 0.73 ).
The bright hotspots are separated by 2.6 arcsec, or a linear size of 22
kpc which means the hotspots may be just within the envelope of the host
galaxy (Spencer \etal 1991). Compact steep spectrum sources and compact
symmetric objects appear to show a high incidence of HI absorption
(Conway 1996).  High resolution radio observations of 3C 190 suggest
one-sided jet-like structure towards the south-west of the core (Spencer
\etal 1991). The large scale jet in this direction is also brighter. The
core is self-absorbed at 608 MHz (Nan Rendong \etal 1991). There appears
to be significant bending between the milli-arcsec scale jet and the kpc
scale jet. The milli-arcsec scale jet is directed towards the west (Hough
et al. 2002), while on larger scales the jet has bent towards south-west
ending with multiple hotspots. The diffuse lobe in this direction is
further towards south-west ($\sim$ 1.5 arcsec) and also misaligned
(Spencer \etal 1991).  This is suggestive of the presence of a dense
inter stellar medium (ISM) causing the jet to bend. Recent detailed
optical spectroscopy of this object showed narrow [O II], [Ne III] and
C III] emission lines, which provide a redshift of 1.1946$\pm$0.0005
(Stockton and Ridgway 2001). In addition, Mg I, Fe II and strong Mg II
absorption has also been detected, but at a slightly higher redshift of
1.19565$\pm$0.00004, corresponding to an infall velocity of 145 \kms ~in
the quasar frame (Stockton and Ridgway 2001). We adopt the redshift of
1.1946$\pm$0.0005 determined from narrow emission lines as the systemic
redshift of the quasar. Its optical to infra-red continuum falls steeply,
causing it to be classified as a "red" quasar (Simpson and Rawlings 2000;
Smith and Spinrad 1980).  The presence of strong Mg II absorption close
to the systemic redshift suggests the presence of intervening gas and
dust, which may also be responsible for reddening the quasar spectrum and
causing the bending of radio jets. Higher chance of detection of \h1a from
3C 190 was foreseen due to two factors: (i) it is a CSS radio source.
Associated optical/UV absorption is most common in CSS quasars among
radio-loud quasars (Baker \etal 2002), and CSS also shows higher chance
of detecting \h1a (Conway 1996). (ii) it is also a "red" quasar which
again shows higher chance of detecting \h1a (C98).

In this paper, we report the detection of \h1a from the red quasar 3C190,
which is blue-shifted with respect to the systemic redshift.  In Section
2 observations and results are presented. The \h1a detected in 3C190 is
discussed in Section 3. Conclusions are presented in Section 4.

\section{Observations and Results}

The observations were carried out using the Giant  Meterwave Radio
Telescope (GMRT) during September 2001. The first set of observations
adopted a bandwidth of 4 MHz centered at 647.5 MHz. Subsequently, two
sets of observations each with a bandwidth of 8 MHz centered at 646.7
MHz and 648.7 MHz were repeated to confirm the detection from the first
set of observations.  The integration time on source was $\sim$ 3 hr in
each set of observations. The velocity resolutions were $\sim$ 29 \kms
~and $\sim$ 14.5 \kms ~for bandwidths of 8 MHz and 4 MHz respectively.
The flux densities were estimated by observing the standard primary
calibrators 3C286 or 3C48. The bandpass calibrators 3C 48, 3C147 and
3C 286 were observed at the beginning as well as at the end of the
observation in order to check for bandpass stability with time. Phase
calibrators were observed for 10 minutes every 30 minutes.

The data obtained from the GMRT were converted to FITS and analysed
using the Astronomical Image Processing System ({\tt AIPS}) following
standard procedures. About 20 line-free channels were collapsed to
obtain a continuum image. A few iterations of phase-only self-calibration
were performed followed by amplitude and phase self-calibration. These
calibrations were then applied to the spectral line data.  The continuum
flux density of 3C 190 at 647 MHz is 5.69 Jy.  The continuum flux density
for subtracting from the spectral data was obtained by averaging several
line-free channels. The final spectral cube was made from this continuum
subtracted data.

Initially, individual spectra were obtained from the two 8 and one 4 MHz
observations to check for consistency.  The individual spectra were then
averaged to obtain the final spectrum which is presented in Figure 1.
The systemic redshift and the relative velocity of the associated optical
absorption system (Stockton and Ridgway 2001) are marked in Figure 1.
The RMS noise in the averaged spectrum is $\sim$ 2 mJy/beam/channel.
We have used the standard {\tt GIPSY} package to fit Gaussians to the
observed spectra. A minimum of 5 Gaussians were required to minimise
the residuals. The estimated parameters of the fit are given in Table 1.

\begin{figure}[t]
\hspace{0.05in}
\psfig{figure=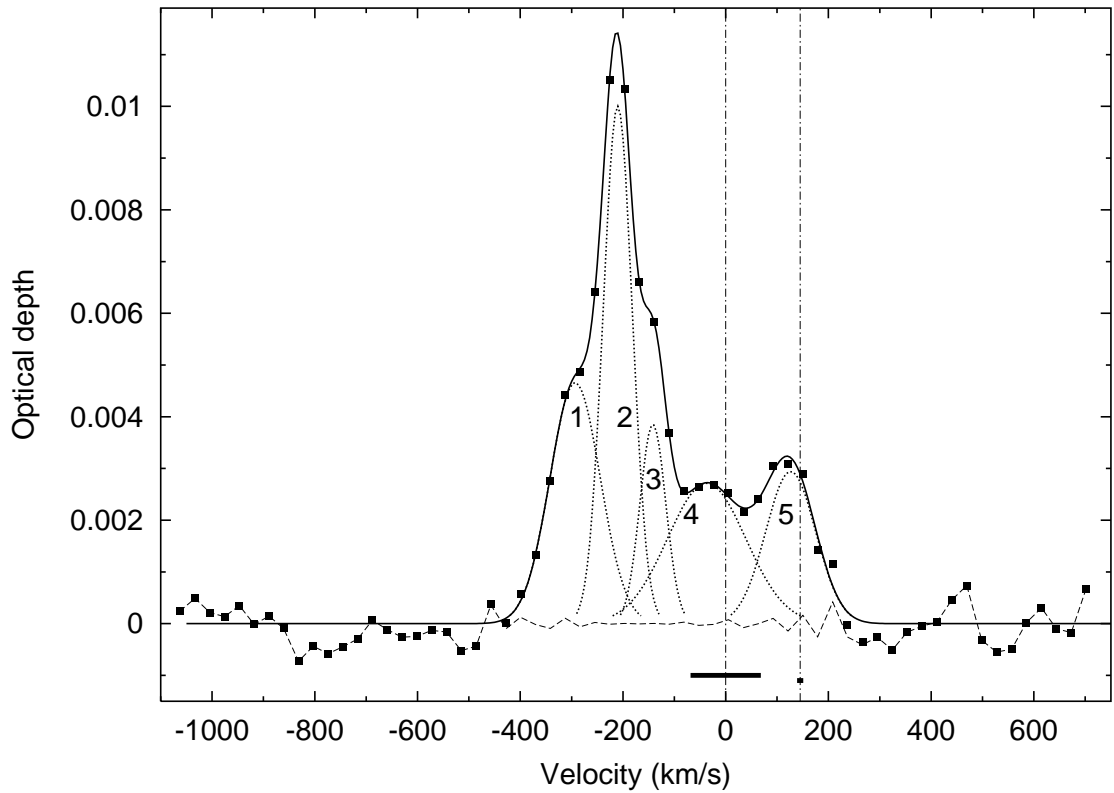,width=15.5cm,angle=-90}
\caption{The HI absorption spectrum toward the red quasar 3C 190
(indicated by filled squares). The solid line is the five component
Gaussian fit to the data and the dotted lines are the individual Gaussian
components which are numbered. Dashed line is the residual to the fit.
The optical depth was estimated using the peak flux of 5.69 Jy.
The systemic redshift is taken as 1.1946$\pm$0.0005 obtained from
narrow [O II], [Ne III]  and C III] emission lines, which is indicated
as dot-dashed vertical line at the velocity of 0 \kms. Similar vertical
line at a velocity of 145 \kms ~corresponds to the Mg II absorption. Thick
horizontal lines centered at 0 and 145  \kms ~indicates one sigma range
of velocity at systemic redshift and that at Mg II absorption.
}

\end{figure}

The most interesting finding from our observations is that most of the
\h1a in 3C 190 is blue-shifted with respect to the systemic redshift
(Figure 1). The peak of the absorption occurs at a velocity of $-$210
\kms, but the overall absorption is very broad, extending from $-400$
\kms ~to +200 \kms. The velocity uncertainties in systemic redshift is 68
\kms ~and that of Mg II absorption system is 5.5 \kms.  The components
1, 2 and 3 are significantly blue-shifted with respect to the systemic
redshift, while component 4 is within the uncertainty in the systemic
redshift.  Component 5 at 127 \kms ~is outside the uncertainty in the
Mg II absorption system and the uncertainty in the systemic redshift.
The three blue-shifted lines also have larger optical depths than lines 4
and 5.  Component 1 at -293 \kms ~has an optical depth of $\sim$ 0.005 and
the second component at -210 \kms ~has an optical depth of $\sim$ 0.01.
The optical depth of component 3 is comparable to that of component 1, but
FWHM is nearly half of component 1.  Components 2 and 3 have comparable
FWHM, but optical depth of component 3 is less than half of component
1. The FWHM of fifth component is comparable to component 1, with optical
depth nearly half. Component 4 is well within uncertainty in the systemic
redshift, and is broadest of all components with FWHM of 179 \kms.

\section{Discussion}

A few cases of blue-shifted \h1a with respect to the systemic redshift are
reported in the literature. A nearby CSS (z = 0.06393) PKS B1814$-$63,
exhibits an HI absorption spectrum similar to that seen in 3C 190, viz,
a broad shallow feature and blue-shifted deep absorption (Morganti
\etal 2001).  A similar absorption profile is also seen in the radio
galaxy 4C 12.50 (Mirabel 1989) where the overall absorption extends
over 900 \kms.  Another compact radio galaxy, 3C 459, also shows HI
absorption blue-shifted with respect to systemic redshift, though
weaker than PKS 1814$-$63 (Morganti \etal 2001). The authors suggest
that the blue-shift could be due to outflow, possibly associated with
the jet/lobes, and the shallow component seen at the systemic redshift
is due to a circumnuclear disk. Significantly blue-shifted \h1a has
also been seen in several Seyfert galaxies. In IC 5063, the \h1a is
blue-shifted by more than 600 \kms ~(Oosterloo \etal 2000) with respect
to the systemic redshift. They suggest that the \h1a could be occurring
from the gas which has been shocked by the western radio jet. The moderate
blue-shift of \h1a observed in NGC 1068 and NGC 3079 is suggested to be
due to neutral gas participating in the outflow (Gallimore \etal 1994).

\begin{table}
\caption{Parameters derived from the Gaussian fitting to the optical
depth profile towards 3C 190. Zero velocity corresponds to the systemic
redshift of 1.1946. }

\begin{tabular}{l l l c}
          &                  &             &            \\
Line      & Velocity         & FWHM  & $\tau$     \\
          & (\kms)           & (\kms)      &            \\
Comp  1  &$-$293.1 $\pm$ 2.8 &115.1 $\pm$ 6.9 & 0.0047 $\pm$ 0.0002 \\
Comp  2  &$-$210.2 $\pm$ 1.0 & ~66.8 $\pm$ 2.2 & 0.0100 $\pm$ 0.0003 \\
Comp  3  &$-$141.8 $\pm$ 2.2 & ~59.4 $\pm$ 5.0 & 0.0043 $\pm$ 0.0003 \\
Comp  4  &$-$37.1 $\pm$ 5.7  &179.4 $\pm$ 13.3& 0.0027 $\pm$ 0.0002 \\
Comp  5  & ~~126.5 $\pm$ 4.2   &114.6 $\pm$ 9.8 & 0.0029 $\pm$ 0.0002 \\
\end{tabular}
\end{table}

In the case of 3C 190, the striking aspect is that the \h1a is broad
and complex and most of the absorption is blue-shifted with respect
to systemic redshift (Figure 1). The total velocity extent is about
600 \kms. The absorption is not against the core because the core is
self-absorbed at this frequency. The flux limit for the core is $<$ 4 mJy
(Nan Rendong \etal 1991), whereas the absorbed flux is up to 70 mJy.
In the current observations, 3C 190 remains unresolved. In the arcsec
resolution map of 3C 190, there are two hotspots in the south-west
direction, and are brighter than the north-east hotspot (Spencer \etal
1991). The fluxes of individual hotspots are about a Jy and more at 608
MHz (Nan Rendong \etal 1991).  The total projected extent of hotspots in
the south-west and north-east is about 22 kpc, hence these hotspots may
lie within the envelope of the host galaxy environment. At this linear
scale, there is also evidence for diffuse radio emission (Pearson \etal
1985) which is mostly north of the radio core. Significant \h1a can
occur against one or all of these. The diffuse lobe in the south west
is further away (1.5 arcsec) from bright hotspots (Pearson \etal 1985),
and may lie outside the host galaxy, hence it is less likely to expect
\h1a against this diffuse lobe.

Even though it is not possible to say which absorption-line is towards
which radio component, we can draw indirect conclusions from the
known radio and optical morphologies and properties. From the radio
image at arc-sec resolution, we see that the south-west diffuse lobe
and the farthest hotspot in this direction are mis-aligned (Spencer
\etal 1991).  The immediate surroundings of 3C 190 have been studied in
imaging as well as in spectroscopy in great detail using HST and Keck
(Stockton and Ridgway 2001).  The most interesting feature revealed
from the HST observations is a linear knotty optical filament along
north-west, very close to the quasar and overlapping (in projection)
with the south-west hotspot.  Spectroscopy of the linear filament reveals
interesting features. The [O II] line profile clearly divides into low
velocity dispersion ($\sigma \sim 40$ \kms) and high velocity dispersion
($\sigma \sim 200$ \kms) groups. The higher velocity dispersion indicates
that the gas might have been shocked in some way by the interaction
with the jet thereby increasing the turbulent velocity in this region
significantly. In the radio map also, there is evidence for jet bending
in this region (Stockton and Ridgway 2001), which indicates the presence
of a dense medium.

Thus it appears that the overall environment under which the \h1a is
occurring in 3C 190 appears to be similar to that of IC 5063, where
neutral gas has been shocked by the radio jet.  Velocities up to 500 \kms
~are expected from shocks (Dopita \& Sutherland 1995).  Due to shocks,
the neutral gas could acquire significant bulk motion but the internal
velocities might not increase significantly.  Such clouds could produce
reasonably narrow lines (FWHM $\sim$ 100 \kms ~or less), but significantly
blue-shifted, such as the lines 1, 2 and 3 in 3C 190 (Table 1), where the
FWHM is 115, 67 and 59 \kms ~respectively, but are blue-shifted by 293,
210 and 142 \kms ~with respect to the systemic redshift.

The optical linear filament which shows high velocity dispersion gas
overlaps (in projection) with the south-west hotspot (Stockton and
Ridgway 2001).  This is suggestive of interaction of the jet with the
dense ambient medium responsible for changing the direction of the
jet. Thus we can expect that the three blue-shifted absorption-lines
(1, 2 and 3) in 3C 190 could be arising from this region. Only one of
the absorption-line component (component 4) in 3C 190 is very broad
(FWHM of 179 \kms). This component is well within the uncertainty in
the systemic redshift and could be due to more turbulent gas near the
region of interaction of radio jet with the ambient medium.  Thus, we
suggest that lines 1, 2, 3 and 4 are likely to be against the south-west
hotspots. The fifth line has a velocity of 127 \kms, which is outside
the uncertainty of Mg II absorption system and the uncertainty in the
systemic redshift. Since the FWHM of this line is not too narrow (115 \kms),
it could be from the ISM of the host galaxy.

\section{Conclusions}

We have reported the GMRT detection of HI 21 cm-line absorption in
the z=1.1946 red quasar 3C 190.  The absorption, with peak at $\sim$
647.7 MHz, is broad and complex, spanning a velocity width of $\sim$
600 \kms. Most of the absorption is blue-shifted with respect to the
systemic redshift. 3C 190 is unresolved with the present observations.
Since the core is self-absorbed at this frequency the absorption must be
towards the hotspots. Comparison of the radio and deep optical images
reveal linear filaments in the optical which overlap with the brighter
radio jet in the south-west direction.  We therefore suggest that the
blue-shifted \h1a could be occurring in the atomic gas shocked by the
south-west jet.

\section*{Acknowledgments}

We thank Rajaram Nityananda, Jayaram Chengalur and Chris Carilli for
comments on the manuscript. We thank the referee for meticulously
reading the manuscript and for constructive suggestions.  We thank the
staff of the GMRT that made these observations possible. GMRT is run
by the National Center for Radio Astrophysics of the Tata Institute
of Fundamental Research.  This research has made use of the NASA/IPAC
Extragalactic Database (NED) which is operated by the Jet Propulsion
Laboratory, California Institute of Technology, under contract with the
National Aeronautics and Space Administration (NASA).

\end{document}